\documentclass{optica-article}
\journal{opticajournal} 
\articletype{Research Article}
\usepackage{lineno}
\usepackage{tabularx}
\usepackage{array} 

\begin{document}

\title{On-Demand Control of Input-State-Dependent Single-Photon Scattering in Multi-Mode Waveguides}
\author{Yan Liu,\authormark{1} Qing-Ao Xiang,\authormark{1} Xin-Yuan Yang,\authormark{1} Ji-Bing Yuan,\authormark{1} Shi-Qing Tang,\authormark{1} Xin-Wen Wang,\authormark{1} and Ya-Ju Song\authormark{1,*}}

\address{\authormark{1}Key Laboratory of Opto-electronic Control and Detection Technology of University of Hunan Province, and College of Physics and Electronic Engineering, Hengyang Normal University, Hengyang 421002, China}

\email{\authormark{*}yjsong@hynu.edu.cn} 

\begin{abstract*} 
Precise control of a single photon transport in broadband, multi-mode waveguides is a fundamental challenge for scalable quantum networks. We propose a theoretical scheme for on-demand control of single-photon scattering using a driven $\Lambda$-type emitter coupled to a rectangular waveguide. By employing the Lippmann-Schwinger formalism, we derive the exact analytical scattering matrix and reveal two key interference mechanisms: electromagnetically induced transparency for complete transmission and Fano resonance for complete reflection. We demonstrate that the single-photon scattering is dynamically engineered by the driving field, enabling a switch between complete transmission and  dual-frequency complete reflection. Crucially, in the multi-mode regime, we show that the scattering is governed by quantum interference between modes, making it critically dependent on the input photonic state. By preparing the photon in a specific coherent superposition state, the multi-mode interference is harnessed to achieve Fano resonance-mediated complete reflection. Conversely, a single-mode input suppresses complete reflection. This input-state-dependent scattering establishes a general framework for multi-mode quantum photonics, paving the way for broadband dual-frequency filters, multi-mode quantum routers, and on-chip spectrometers.
\end{abstract*}

\section{\label{sec:level1}Introduction}
Waveguide quantum electrodynamics (waveguide QED) investigates the interaction between quantum emitters (e.g., atoms, quantum dots, or superconducting qubits) and photons confined in waveguides \cite{GU20171,gonzalez2024light,PhysRevLett.95.213001,PhysRevLett.98.153003,RevModPhys.89.021001}. 
Compared to free-space or cavity QED, waveguide QED offers enhanced scalability and integrability, enabling applications such as high-fidelity quantum gates \cite{Paulisch_2016}, deterministic quantum state transfer, and scalable quantum networks \cite{Kimble2008,PhysRevLett.101.100501,PhysRevLett.111.103604}. 
Precise control of single-photon transport is essential for these quantum functionalities \cite{Bharath2023} and is typically achieved via quantum interference. For example, electromagnetically induced transparency (EIT) creates a narrow transparency window with low absorption \cite{PhysRevA.108.063702}, whereas the Fano resonance arises from interference between a discrete state and a continuum, yielding a characteristic asymmetric spectral line shape \cite{PhysRev.124.1866,PhysRevA.89.063810}. These fundamental mechanisms have been further enriched by incorporating principles from non-Hermitian physics \cite{PhysRevA.99.063806}, giant atom systems \cite{Bharath2020,Sun_2023,Zhang2023,Zhou_2024,PhysRevA.111.033701,Jia2024,Sun2025}, topological waveguides \cite{PhysRevX.11.011015,PhysRevA.111.023711}, and non-perturbative theories \cite{PhysRevResearch.4.023194}.

However, most theoretical studies of waveguide QED idealize photons as confined to a single fundamental mode. This simplification  reduces the photon-emitter interaction to a tractable one-dimensional (1D) scattering problem \cite{PhysRevA.108.043709,PhysRevA.109.063708,PhysRevA.108.053718,Yan2015}, often facilitated by approximations such as Wigner-Weisskopf approximation and linear dispersion model. In reality, however, quantum photonic platforms, including hollow-core fibers \cite{Mitsch2014,PhysRevX.5.041036}, photonic crystal waveguides \cite{PhysRevLett.113.113904,Mahmoodian2017}, and superconducting circuits \cite{Bharath2023,PhysRevX.13.021039,PRXQuantum.5.030346}, can be engineered to support multiple propagation modes, leading to modal dispersion. While this multi-mode capability offers advantages for information capacity expansion and high-dimensional quantum encoding \cite{PhysRevA.70.042313}, it introduces significant challenges for deterministic photon transport due to inter-mode interference and unwanted scattering into non-target channels. This issue becomes particularly acute in emerging high-frequency applications, such as X-rays and free-electron lasers \cite{Spiller1974,PhysRevLett.109.233907,PhysRevA.109.033703}, where the photon wavelength can be comparable to or shorter than the waveguide’s transverse dimensions. In such regimes, the coexistence of numerous transverse electric (TE) and magnetic (TM) modes \cite{PhysRevA.88.013836,Song_2018,Zeng_2023,Song_2024,Song:20} transforms photon dynamics into a complex multi-channel problem, underscoring the urgent need for theoretical frameworks that can harness or mitigate multi-mode effects in waveguide QED.

In this paper, we theoretically propose a scheme for input-state-dependent single-photon transport in a rectangular waveguide coupled to a driven $\Lambda$-type quantum emitter. We consider a configuration where one transition of the $\Lambda$-emitter is directly coupled to the waveguide, while the second transition is coherently driven by an external field \cite{Zhang_2024,Witthaut_2010,Martens_2013,dkbz-thfh,PhysRevA.109.033710}.  By employing the Lippmann-Schwinger formalism \cite{PhysRevA.93.013828,Taylor,Tannoudji}, we derive an exact analytical scattering matrix valid for both single- and multi-mode regimes. Our analysis reveals two distinct quantum interference mechanisms: EIT enabling complete transmission and Fano resonance yielding complete reflection. Crucially, we demonstrate that in the multi-mode regime, the scattering behavior exhibits sensitivity to the input photonic state—complete reflection is exclusively achieved for specific coherent superposition states, whereas single-mode inputs inherently suppress complete reflection. This work provides an exact framework for multi-mode waveguide QED and establishes a generalized toolbox for multi-mode quantum photonics, paving the way for broadband dual-frequency filtering, mode-selective routing, and on-chip spectrometry.

The paper is structured as follows: Section~\ref{sec2} presents the physical model and derives the multi-mode scattering formalism. Section~\ref{sec3} establishes the exact analytical conditions for complete transmission and reflection in multi-mode waveguide QED systems, investigating the modulation mechanism of the driving field within a dressed-state framework. Section~\ref{sec4} investigates the precise control of single-photon transport, enabled by a driven $\Lambda$-type emitter, in both single- and multi-mode regimes. Finally, conclusions are drawn in Section~\ref{sec5}.

\section{Physical model and multi-mode scattering formalism}\label{sec2}
In this section, we first introduce the physical model and Hamiltonian describing an infinite rectangular waveguide coupled to a driven $\Lambda$-type emitter. We then derive the exact multi-mode scattering matrix for this waveguide QED system using the Lippmann-Schwinger formalism, obtaining closed-form expressions for the reflection and transmission amplitudes of arbitrary single-photon input states.

\begin{figure}[ht]
\centering
\includegraphics[width=8cm]{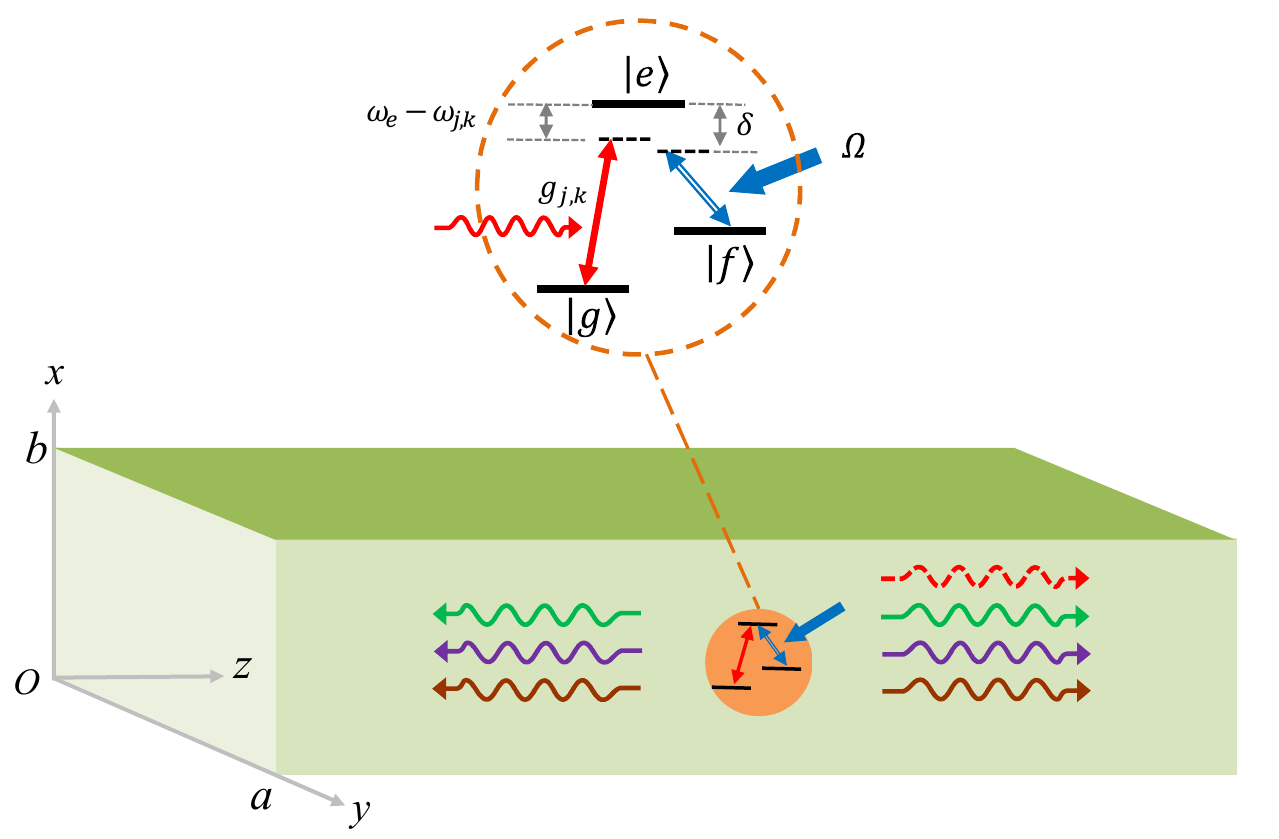}
\caption{Schematic of a driven $\Lambda$-type emitter coupled to an infinite rectangular waveguide (cross-section $A=ab, a=1.5b$). The waveguide couples $\vert e\rangle\leftrightarrow\vert g\rangle$ with strength $g_{j,k}$ and detuning $\omega_{e}-\omega_{j,k}$, while a classical field drives $\vert e\rangle\leftrightarrow\vert f\rangle$ with Rabi frequency $\Omega$ and detuning $\delta$.
}
\label{model}
\end{figure}

We consider a driven $\Lambda$-type three-level emitter coupled to an infinitely long rectangular waveguide with cross-sectional dimensions $a=1.5b$ (see Fig.~\ref{model}). The emitter, located at position $(a/2,b/2,z_{0})$, consists of a ground state $\vert g\rangle$, an intermediate state $\vert f\rangle$, and an excited state $\vert e\rangle$, with respective energies $\omega_{g}$, $\omega_{f}$, and $\omega_{e}$. Here, $\omega_{g}$ is set to zero as a reference. The emitter's $z$-oriented dipole moment selectively interacts with the TM modes. The waveguide supports TM$_{mn}$ modes indexed by $j=1,2,3,\dots$ corresponding to TM$_{11}$, TM$_{31}$, TM$_{13}$, etc., ordered by increasing cutoff frequencies $\omega_{j}= \pi\sqrt{m^{2}/a^{2}+n^{2}/b^{2}}$ (in units where $\hbar=c=1$). Each mode is described by annihilation operators $\hat{a}_{j,k}$ and creation operators $\hat{a}^{\dagger}_{j,k}$ for photons of wavevector $\mathbf{k}=(m\pi/a,n\pi/b, k)$ with $k$ the longitudinal wave number. The waveguide obeys the dispersion relation $\omega_{j,k}=\sqrt{\omega_{j}^{2}+k^{2}}$. The waveguide TM modes mediate the $\vert e \rangle \leftrightarrow \vert g\rangle$ transition, whereas a classical field characterized by Rabi frequency $\Omega$ and detuning $\delta$ coherently drives the $\vert e \rangle \leftrightarrow \vert f\rangle$ transition. 

Within the rotating-wave approximation and in a rotating frame with respect to $\hat{H_{0}^{\prime}}=(\omega_{f}-\omega_{e}+\delta)\vert f\rangle\langle f\lvert$, the system Hamiltonian $\hat{H}=\hat{H}_{0}+\hat{V}$ comprises \cite{PhysRevA.89.063810,PhysRevA.88.013836,Song_2018,Zeng_2023}:
\begin{gather}
\hat{H}_{0}=\omega _{e}\vert e\rangle \langle e\vert
+(\omega _{e}-\delta) \vert f\rangle \langle f\vert +\sum_{j}\int_{-\infty }^{+\infty }dk\,\omega_{j,k}\hat{a}_{j,k}^{\dagger}\hat{a}_{j,k},\label{eq-1} \\
\hat{V}=\sum_{j}\int_{-\infty }^{+\infty }dk\,( g_{j,k}\vert
e\rangle \langle g\vert \hat{a}_{j,k}+ g_{j,k}^{*}\vert
g\rangle \langle e\vert \hat{a}^{\dagger}_{j,k})+\Omega (\vert e\rangle \langle f\vert +\vert f\rangle \langle e\vert). \label{eq-2}
\end{gather}
where the free Hamiltonian $\hat{H}_{0}$ describes the uncoupled emitter and waveguide modes, and the interaction Hamiltonian $\hat{V}$ contains both emitter-waveguide coupling and classical field driving. The emitter-waveguide coupling strength is given by $g_{j,k}=-g\omega_{j}\sin(m\pi/2)\sin(n\pi/2)e^{-ikz_{0}}/\sqrt{\omega_{j,k}}$, where $g=\mu_{e}/\sqrt{\varepsilon\pi A}$ is determined by the dipole matrix element $\mu_{e}$ and the cross-sectional area $A= ab$ \cite{PhysRevA.89.063810}. The coupling strength vanishes for modes with even $m$ or $n$.

To characterize the single-photon scattering process, we introduce the scattering matrix $S$ that relates the asymptotic input and output states. The system is initialized with the emitter in its ground state $\vert g\rangle$ and a single photon in the TM$_{j}$ waveguide mode $\vert\varphi_{j,k}\rangle =\hat{a}^{\dagger}_{j,k}\vert 0\rangle$, where $\vert 0\rangle$ denotes the photonic vacuum and $k$ is the longitudinal wave number. The input state $\vert\varphi_{j,k}\rangle\vert g\rangle$ is an eigenstate of the free Hamiltonian $\hat{H}_{0}$ with energy $\omega=\omega_{j,k}$. The scattering state $\vert \varphi _{j,k}^{(+)}\rangle$ is an eigenstate of the total Hamiltonian $\hat{H}=\hat{H}_{0}+\hat{V}$ at the same energy and satisfies the Lippmann-Schwinger equation \cite{Taylor}
\begin{equation}
\label{eq-3}
\vert \varphi _{j,k}^{(+)}\rangle = \vert \varphi
_{j,k}\rangle\vert g\rangle +\frac{1}{\omega-\hat{H}_{0}+i0^{+}}\hat{V}%
\vert \varphi _{j,k}^{(+)}\rangle.
\end{equation}
According to the principle of excitation-number conservation, the single-excitation scattering state is constrained to the following form 
\begin{equation}
\label{eq-4}
\vert \varphi _{j,k}^{(+)}\rangle = \sum_{j^{\prime }}\int_{-\infty }^{+\infty}
dk^{\prime }u_{j^{\prime },k^{\prime };j,k}\vert \varphi _{j^{\prime
},k^{\prime }}\rangle\vert g\rangle+\alpha _{j,k}\vert 0\rangle\vert e\rangle +\beta
_{j,k}\vert 0\rangle\vert f\rangle,
\end{equation} 
where $u_{j^{\prime },k^{\prime};j,k}$ represents the probability amplitude for photon emission into the TM$_{j^{\prime}}$ mode with longitudinal wave number $k^{\prime}$, while $\alpha_{j,k}$ and $\beta_{j,k}$ denote the excitation amplitudes of the emitter in states $\vert e\rangle$ and $\vert f\rangle$, respectively. Substituting Eq.~(\ref{eq-4}) into Eq.~(\ref{eq-3}) yields the relationship among $u_{j^{\prime },k^{\prime};j,k}$, $\alpha_{j,k}$ and $\beta_{j,k}$. By expressing $u_{j^{\prime },k^{\prime};j,k}$ and $\beta_{j,k}$ in terms of $\alpha_{j,k}$ we obtain the expression for $\alpha_{j,k}$,
\begin{equation}
\label{eq-6}
\alpha _{j,k}=\frac{g_{j,k}}{\omega-\omega _{e}-\frac{\Omega ^{2}}{\omega-\omega_{e}+\delta}-\Delta(\omega)+i\Gamma(\omega)}.
\end{equation}
During the calculation, we used the Sokhotski-Plemelj identity $\lim_{\varepsilon \rightarrow 0^{+}}(x-x_{0}+i\varepsilon)^{-1}=\mathcal{P}[(x-x_{0})^{-1}]-i\pi \delta (x-x_{0})$ (with $\mathcal{P}$ denoting the Cauchy principal value) to decompose the self-energy $\sum (\omega)=\sum_{j}\int_{-\infty }^{+\infty} dk\vert g_{j,k}\vert ^{2}/(\omega-\omega_{j,k}+i0^{+})$ into the Lamb shift $\Delta(\omega)$ and decay rate $\Gamma(\omega)$,
\begin{equation}
\Delta(\omega)=\sum_{j}\Delta_{j}(\omega), \quad
\Gamma(\omega)=\sum_{j}\Gamma_{j}(\omega).\label{eq-7}
\end{equation}
The mode-resolved contributions are defined by
\begin{equation}
\Delta_{j}(\omega)=\mathcal{P}\int_{-\infty }^{+\infty}dk\frac{\vert g_{j,k}\vert ^{2}}{\omega-\omega _{j,k}}, \quad
\Gamma_{j}(\omega)=2\pi\vert g_{j,k_{j}}\vert ^{2}\rho_{j}( \omega),\label{eq-8}
\end{equation}
where the density of states is derived from the dispersion relation
\begin{equation}
\label{eq-9}
\rho_{j}(\omega)=\left|\frac{1}{d\omega/dk|_{k=\pm k_{j}}}\right|=\frac{\omega}{\sqrt{\omega^{2}-\omega_{j}^{2}}}\Theta (\omega-\omega_{j}),
\end{equation}
with $k_{j}=\sqrt{\omega^{2}-\omega_{j}^{2}}$ and $\Theta (\omega-\omega_{j})$ the Heaviside step function. With the interaction Hamiltonian $\hat{V}$ from Eq.~(\ref{eq-2}) and the scattering state from Eq.~(\ref{eq-4}), the on-shell $T$ matrix element is \cite{Taylor}
\begin{equation}
\label{eq-10}
T_{j^{\prime }k^{\prime }\leftarrow jk}=\langle g\vert \langle \varphi _{j^{\prime},k^{\prime }}\vert \hat{V}\vert \varphi_{j,k}^{(+)}\rangle =\alpha _{j,k}g_{j^{\prime },k^{\prime }}^{\ast}.
\end{equation}
Then the scattering operator $\hat{S}$ follows by inserting Eqs.~(\ref{eq-6}) and (\ref{eq-10}) into $S_{j^{\prime }k^{\prime }\leftarrow jk}=\delta _{j,j^{\prime }}\delta
_{k,k^{\prime }}-2\pi i\delta ( \omega _{j^{\prime },k^{\prime
}}-\omega) T_{j^{\prime }k^{\prime }\leftarrow jk}$, resulting in
\begin{equation}
\label{eq-11}
S_{j^{\prime }k^{\prime }\leftarrow jk}=\delta _{j,j^{\prime }}\delta
_{k,k^{\prime }}-\frac{2\pi i g_{j,k}g_{j^{\prime },k^{\prime }}^{\ast }}{G( \omega)}\delta (\omega_{j^{\prime },k^{\prime
}}-\omega),
\end{equation}
where the function $G(\omega)$ is defined as
\begin{equation}
\label{eq-12}
G(\omega) =\omega-\omega _{e}-\frac{\Omega ^{2}}{
\omega-\omega _{e}+\delta }-\Delta( \omega)
+i\Gamma ( \omega).
\end{equation}
Equation~(\ref{eq-11}) describes the single-photon scattering dynamics, where the first term represents direct transmission and the second term describes the emitter-mediated scattering. For an incident photon in the $k$-th TM$_{j}$ mode, the coupling strength $g_{j,k}$ drives the emitter transition $\vert g\rangle\rightarrow\vert e\rangle$. Subsequent decay back to $\vert g\rangle$ enables re-emission into either the original $k$-th TM$_{j}$ mode or a different $k^{\prime}$-th TM$_{j^{\prime}}$ mode, with respective coupling strength $g_{j^{\prime},k^{\prime}}$. This scattering process is characterized by two key features. First, the classical drive applied to the $\vert e\rangle\leftrightarrow\vert f\rangle$ transition  modulates the scattering via the function $G(\omega)$ in Eq.~(\ref{eq-12}). Second, this external field enables interconversion among guided TM modes, which is strictly constrained by energy conservation as dictated by the term $\delta(\omega_{j^{\prime},k^{\prime}}-\omega)$.

For the emitter initially prepared in the ground state $\vert g\rangle$ and an incident photon at frequency $\omega$, the input state can be represented as a superposition over all accessible modes:
\begin{eqnarray}
\label{eq-13}
\vert \varphi_{\text{in}}\rangle &=& \sum_{j = 1}^{j_{\max }}c_{j}\vert \varphi_{j,k_{j}}\rangle,
\end{eqnarray}
where $c_{j}$ are the normalized amplitudes, and the longitudinal wave number of the $j$-th mode is $k_{j}=\sqrt{\omega^{2}-\omega_{j}^{2}}$. The total number of propagating modes, $j_{\max}$, is determined  by the relationship between the photon frequency and the cutoff frequencies, satisfying $\omega_{j_{\max}}<\omega <\omega_{j_{\max}+1}$. While single-mode operation occurs in the range $\omega_{1}<\omega<\omega_{2}$ ($j_{\max}=1$), practical waveguides support multiple TM modes for $\omega >\omega _{2}$ ($j_{\max}>1$). Then the outgoing state is derived using the multi-mode $S$-matrix as described in Eq.~(\ref{eq-11}):
\begin{equation}
\label{eq-14}
\vert \varphi_{\text{out}}\rangle=\hat{S}\vert \varphi_{\text{in}}\rangle =\sum_{j^{\prime } = 1}^{j_{\max }^{\prime }}\sum_{j = 1}^{j_{\max }}\int dk^{\prime}c_{j} S_{j^{\prime }k^{\prime }\leftarrow jk_{j}} \vert\varphi _{j^{\prime},k^{\prime }}\rangle =\sum_{j = 1}^{j_{\max }}
r_{j}\vert
\varphi_{j,-k_{j}}\rangle+t_{j}\vert \varphi _{j,k_{j}}\rangle.
\end{equation}
Here, the reflection and transmission amplitudes are expressed as
\begin{equation}
r_{j} =\frac{-2\pi i\rho_{j}(\omega)
g_{j,-k_{j}}^{\ast }}{G(\omega)}\sum_{j= 1}^{j_{\max }}c_{j}g_{j,k_{j}}, \quad
t_{j} = c_{j}-\frac{2\pi i\rho_{j}(\omega) g_{j,k_{j}}^{\ast}}{G(\omega)}\sum_{j= 1}^{j_{\max }}c_{j}g_{j,k_{j}}.\label{eq-15}
\end{equation}
The reflectance and transmittance in the $j$-th scattering channel are determined from the scattering amplitudes and the group velocities $\vert v_{g,j}\vert=1/\rho_{j}(\omega)$ from Eq.~(\ref{eq-9}), yielding
\begin{equation} 
R_{j}=\frac{\frac{\vert r_{j} \vert^{2}}{\rho_{j}(\omega)}}{\sum_{j=1}^{j_{\max}} \frac{\vert c_{j}\vert^{2}}{\rho_{j}(\omega)}}, \quad 
T_{j} = \frac{\frac{\vert t_{j} \vert^{2}}{\rho_{j}( \omega )}}{\sum_{j=1}^{j_{\max}} \frac{\vert c_{j}\vert^{2}}{\rho_{j}(\omega)}}. 
\label{eq-16} 
\end{equation}
Consequently, the total reflectance $R=\sum_{j=1}^{j_{\max}}R_{j}$ and total transmittance $T=\sum_{j=1}^{j_{\max}}T_{j}$ satisfy
\begin{equation} 
\label{eq-17} R = 1 - T = \frac{2\pi\Gamma(\omega) \left\vert\sum_{j = 1}^{j_{\max}} c_{j} g_{j,k_{j}}\right\vert^{2}}{\vert G (\omega) \vert^{2} \sum_{j = 1}^{j_{\max}} \frac{\vert c_{j}\vert^{2}}{\rho_{j} (\omega)}}. 
\end{equation}

\section{Analytical conditions for complete transmission and reflection}\label{sec3}
This section establishes analytical criteria for complete transmission and reflection in multi-mode waveguide QED systems. Building on the scattering framework of Sec.~\ref{sec2}, we elucidate how EIT enables complete transmission via quantum interference. For the complete reflection analysis, we employ the dressed-states formalism to explain complete reflection induced by Fano resonances. These results provide explicit transport control guidelines applicable to both single- and multi-mode regimes, establishing the foundation for subsequent input-state-dependent protocols.

By substituting Eqs.~(\ref{eq-7}), (\ref{eq-8}), and (\ref{eq-12}) into Eq.~(\ref{eq-17}), we derive the total reflectance as
\begin{equation}
R = \frac{2\pi\Gamma(\omega) \biggl\vert
\sum_{j=1}^{j_{\max}}c_{j}g_{j,k_{j}}\biggr\vert ^{2}}
{\displaystyle \Bigl\{ \operatorname{Re}^{2}\!\bigl[G(\omega)\bigr] + \Gamma^{2}(\omega) \Bigr\} 
\displaystyle \sum_{j=1}^{j_{\max}} \frac{\bigl\vert c_{j}\bigr\vert^{2}}{\rho_{j}(\omega)}}
\leq \frac{\biggl\vert\sum_{j=1}^{j_{\max}}c_{j}g_{j,k_{j}}\biggr\vert ^{2}}
{\displaystyle \sum_{j=1}^{j_{\max}}\bigl\vert g_{j,k_{j}}\bigr\vert^{2} \rho_{j}(\omega)
\displaystyle \sum_{j=1}^{j_{\max }} \frac{\bigl\vert c_{j}\bigr\vert^{2}}{\rho_{j}(\omega)}}
\leq 1.
\label{eq-18}
\end{equation}

Equation~(\ref{eq-18}) demonstrates two mechanisms for achieving complete transmission ($R=0$). First, the dark-state condition, $\sum_{j=1}^{j_{max}}c_{j}g_{j,k_{j}}=0$, suppresses the photon-emitter interaction by satisfying $\hat{V}\vert\varphi_{in}\rangle=0$, which results in perfect transparency. Second, the two-photon resonance condition ($\delta=\omega_{e}-\omega$) induces $\vert G(\omega)\vert\to\infty$, facilitating EIT independently of the input state. In the EIT mechanism, there exist two excitation paths from the $\vert g \rangle$ to the $\vert e \rangle$: one through direct waveguide coupling and the other $\vert g\rangle\rightarrow\vert e\rangle\rightarrow\vert f\rangle\rightarrow\vert e\rangle$ via the intermediate state $\vert f\rangle$. Under the two-photon resonance condition, the probability amplitudes of these two pathways exhibit equal magnitudes but opposite phases, thereby causing destructive interference that suppresses population transfer to $\vert e \rangle$. Consequently, a complete transmission peak (CTP) emerges when the photon frequency $\omega$ is tuned to satisfy $\omega=\omega_{e}-\delta$ for $\delta\in(\omega_{e}-\omega_{j_{\max}+1},\omega_{e}-\omega_{j_{\max}})$, as depicted in the gray-shaded region of Fig.~\ref{fig:SMR}(a). 

After analyzing the transmission characteristics, we investigate the reflection phenomenon. The first inequality in Eq.~(\ref{eq-18}) holds if and only if the real part of $G(\omega)$ vanishes, satisfying the Fano-resonance condition $\text{Re}[G(\omega)]=\omega-\omega_{e}-\Omega^{2}/(\omega-\omega_{e}+\delta)-\Delta(\omega)=0$. By the Cauchy-Schwarz inequality, the second inequality attains equality when either the system operates in the single-mode regime ($j_{\max}=1$), or the photon is prepared in a specific coherent superposition state (SCSS) with $c_{j}\propto\rho_{j}g^{\ast}_{j,k_{j}}$ in the multi-mode regime. The count of complete reflection peak (CRP) is determined by the number of zeros of $\text{Re}[G(\omega)]$ within the frequency band $\omega\in[\omega_{j_{\max}},\omega_{j_{\max}+1})$, which can be precisely controlled by the driving field parameters $\Omega$ and $\delta$. For $\delta$ outside the EIT window $\delta\notin(\omega_{e}-\omega_{j_{\max}+1},\omega_{e}-\omega_{j_{\max}})$, $\text{Re}[G(\omega)]$ is monotonically increasing, yielding at most one CRP. For $\delta$ within the EIT window $\delta\in(\omega_{e}-\omega_{j_{\max}},\omega_{e}-\omega_{j_{\max}+1})$, $\text{Re}[G(\omega)]$ exhibits a divergence at $\omega=\omega_{e}-\delta$ and increases on both sides of this point, yielding zero, one, or two CRPs. As summarized in Table~\ref{tab:ctp_crp}, the CTP count is determined solely by $\delta$, whereas the CRP count depends on $\delta$ and the sign changes of $Re[G(\omega_{j})]$ across the band, the latter being controllable via $\Omega$ and $\delta$. Notably, the derived conditions for complete transmission and reflection do not apply to a photon at the waveguide cutoff frequencies (discussed in Sec.~\ref{sec4}).

\begin{table}[htbp]
\caption{Counts of complete transmission (CTP) and reflection (CRP) peaks$^{a,b}$}
    \label{tab:ctp_crp}
    \centering
\begin{tabular}{|l|l|l|}
     \hline
     \rowcolor[gray]{0.9}
        \text{Detuning Range}&\text{CTP count}&\text{CRP count}\\
\hline
\text{Outside EIT window}&\text{0}&0 if $\text{Re}[G(\omega_{j_{\max}})]\text{Re}[G(\omega_{j_{\max}+1})]>0$ \\
$\delta\notin(\omega_e-\omega_{j_{\max}+1},\omega_e-\omega_{j_{\max}})$&&1 if $\text{Re}[G(\omega_{j_{\max}})]<0<\text{Re}[G(\omega_{j_{\max}+1})]$ \\
\hline
\text{Within EIT window}&1&0 if $\text{Re}[G(\omega_{j_{\max}})]>0>\text{Re}[G(\omega_{j_{\max}+1})]$ \\
$\delta\in(\omega_e-\omega_{j_{\max}+1},\omega_e-\omega_{j_{\max}})$&&1  if $\text{Re}[G(\omega_{j_{\max}})]\text{Re}[G(\omega_{j_{\max}+1})]>0$\\
&&2 if $\text{Re}[G(\omega_{j_{\max}})]<0<\text{Re}[G(\omega_{j_{\max}+1})]$\\
\hline
   \end{tabular}
   $^a$CTP condition:$\delta=\omega_{e}-\omega$; CRP condition:$\text{Re}[G(\omega)]=0$ \text{and} ($j_{\max}=1$ \text{or} $c_{j}\propto\rho_{j}g^{\ast}_{j,k_{j}}$).\\
$^b$ Excluding dark-state and mode cutoffs effects.
\end{table}

To reveal the physics underlying complete-reflection control, we adopt the dressed-state basis. The classical drive diagonalizes the $\{\vert e\rangle,\vert f\rangle\}$ subspace into
\begin{equation}
\vert \nu_{+}\rangle=\sin\theta\vert e\rangle+\cos\theta\vert f\rangle,\quad 
\vert \nu_{-}\rangle=-\cos\theta\vert e\rangle+\sin\theta\vert f\rangle,\label{eq-20}
\end{equation}
with $\tan\theta=2\Omega/(\sqrt{4\Omega^{2}+\delta^{2}}-\delta)$ and eigenenergies $\nu_{\pm}=\omega_{e}+(\delta\pm\sqrt{4\Omega^{2}+\delta^{2}})/2$. The total Hamiltonian can be rewritten as
\begin{equation}
\begin{split}
\hat{H}&=\sum_{i=\pm}\nu_{i}\vert\nu_{i}\rangle\langle\nu_{i}\vert+\sum_{j}\int_{-\infty }^{+\infty }dk\omega_{j,k}\hat{a}_{j,k}^{\dagger}\hat{a}_{j,k}\\
&+\sum_{j}\int_{-\infty }^{+\infty }dk( \sin\theta g_{j,k}\vert\nu_{+}\rangle \langle g\vert \hat{a}_{j,k}-\cos\theta g_{j,k}\vert
\nu_{-}\rangle \langle g\vert \hat{a}_{j,k}+H.c.).
\end{split}
\label{eq-22}
\end{equation}
Thus, the driven $\Lambda$-type emitter behaves as an effective V-type emitter whose transitions $\vert g\rangle\leftrightarrow\vert\nu_{\pm}\rangle$ couple to the waveguide with tunable strengths $\sin\theta g_{j,k}$ and $-\cos\theta g_{j,k}$, respectively. The complete reflection condition $\text{Re}[G(\omega)]=0$ can also be expressed as 
\begin{equation}
[\omega-\nu_{+}-\sin^{2}\theta\Delta(\omega)][\omega-\nu_{-}-\cos^{2}\theta\Delta(\omega)]=\sin^{2}\theta\cos^{2}\theta\Delta^{2}(\omega).\label{eq-23}
\end{equation}
In the weak-coupling limit, we neglect the second-order Lamb shift term and approximate
$\Delta(\omega)\approx\Delta(\omega_{e})$, yielding the renormalized resonance frequencies
\begin{equation}
\tilde{\nu}_{+}=\nu_{+}+\sin^{2}\theta\Delta(\omega_{e}),\quad 
\tilde{\nu}_{-}=\nu_{-}+\cos^{2}\theta\Delta(\omega_{e}).\label{eq-24}
\end{equation}
Figure~\ref{fig:SMR}(b) superimposes the exact zeros of $\text{Re}[G(\omega)]$ (solid black curves) and the approximate conditions $\omega=\tilde{\nu}_{\pm}$ (dotted white curves) in the single-mode regime. Their near-perfect overlap confirms that complete reflection occurs whenever the input photon matches the tunable renormalized transition frequencies $\tilde{\nu}_{\pm}$. This complete reflection in the single-mode regime is induced by the Fano resonance: at $\omega=\tilde{\nu}_{\pm}$ the directly transmitted and re-emitted waves acquire a relative phase of $\pi$ as described in Eq.~(\ref{eq-15}), leading to vanishing transmission. 

\section{Numerical demonstration of input-state-dependent photon transport control}\label{sec4}
In this section, we numerically investigate the precise modulation of single-photon transport by the driven $\Lambda$-type emitter in both single- and multi-mode regimes, and analyze the dependence of this control on the incident photonic state.

\subsection{Single-mode regime: precision tuning via the driven $\Lambda$-type emitter}
Single-mode operation is defined by the spectral window $\omega_{1}<\omega <\omega_{2}$, in which only the fundamental TM$_{11}$ mode ($j=1$) propagates. All frequencies are normalized to $c/b$, where $c$ is the speed of light and $b$ is the shorter transverse dimension of the rectangular waveguide. For $a = 1.5b$, the normalized cutoff frequencies are $\omega_{1}=3.78$ and $\omega_{2}=7.02$. 

\begin{figure}[ht]
\centering
\includegraphics[clip=true,width=13cm]{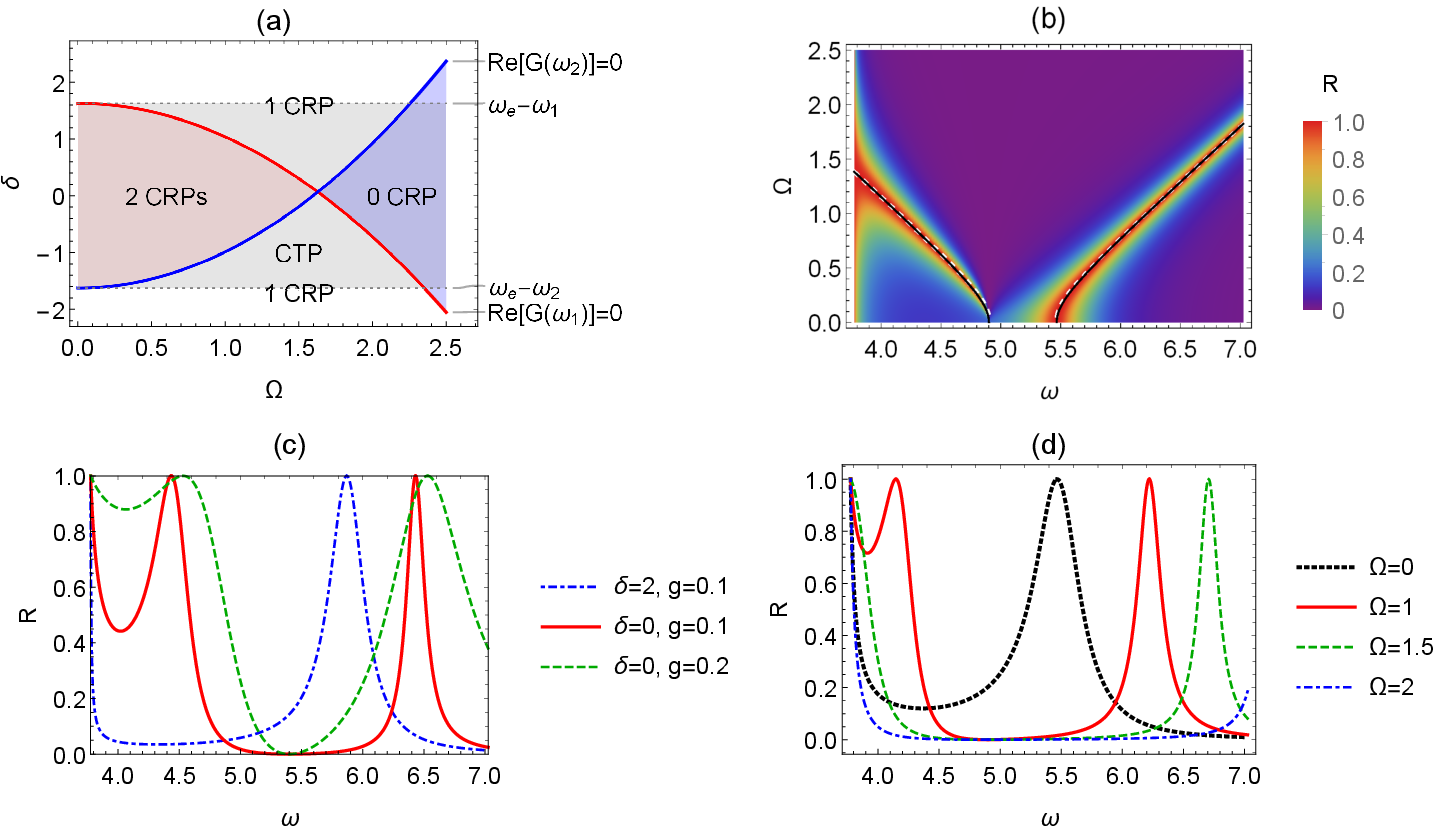}
\caption{Single-mode regime analysis. (a) Phase map showing complete reflection peak (CRP) and complete transmission peak (CTP) counts across varied driving field parameters $\Omega$ and $\delta$; (b) Reflectance $R$ as a function of $\omega$ and $\Omega$, where black thick curves indicate complete reflection at $\text{Re}[G(\omega)]=0$, and white dotted curves represent $\omega=\tilde{\nu}_{\pm}$; (c) $R$ versus $\omega$ for $(\delta,g)=(2,0.1),(0,0.1),(0,0.2)$ at $\Omega=1$ ; (d) $R$ versus $\omega$ for $\Omega=0,1,1.5,2$. In (b) and (d), $g=0.1$ and $\delta=0.5$. Other parameters include $a=1.5b$, with cutoff frequencies $\omega_{1}=3.78$, $\omega_{2}=7.02$, and $\omega_{e}=(\omega_{1}+\omega_{2})/2$. Frequencies are normalized to $c/b$.}
\label{fig:SMR}
\end{figure}

The emitter is initialized in the $\vert g\rangle$, and the incident photon is in the single-photon state $\vert\varphi_{in}\rangle=\vert \varphi_{1,k_{1}}\rangle$ with frequency $\omega$ and longitudinal wave number $k_{1}=\sqrt{\omega^{2}-\omega_{1}^{2}}$. By restricting the system to single-mode operation ($j_{\max}=1, c_{j}=\delta_{j,1}$), the reflectance in Eq.~(\ref{eq-17}) becomes
\begin{equation}
\label{eq-25}
R=1-T=\frac{\Gamma _{1}^{2}\left( \omega \right)}{\left\vert G\left( \omega
\right) \right\vert ^{2}},
\end{equation}
where $G(\omega)=\omega-\omega_{e}-\Omega ^{2}/(\omega-\omega _{e}+\delta)-\Delta_{1}(\omega)+i\Gamma_{1}( \omega)$. 

Complete transmission ($R=0$) occurs at $\omega=\omega_{e}-\delta$. When the drive detuning $\delta$ within the EIT window $\omega_{e}-\omega_{2}<\delta<\omega_{e}-\omega_{1}$, the system exhibits a single CTP within the single-mode band, as shown by the gray-shaded region in Fig.~\ref{fig:SMR}(a) and the solid red and dashed green curves in Fig.~\ref{fig:SMR}(c). Conversely, for $\delta$ outside this range, complete transparency is suppressed (dot-dashed blue curve in Fig.~\ref{fig:SMR}(c)). 

Equation~(\ref{eq-25}) reveals two conditions for complete reflection. First, at the cutoff frequency $\omega=\omega_{1}$, the group velocity $v_{g}=(1/\rho_{1})\vert_{\omega=\omega_{1}}=0$, resulting in complete reflection. Second, complete reflection occurs when the Fano resonance condition $\text{Re}[G(\omega)]=\omega-\omega_{e}-\Omega ^{2}/(\omega-\omega _{e}+\delta)-\Delta_{1}(\omega)=0$ is satisfied. 
In this case, the reflection spectrum is highly sensitive to the driving field. Fig.~\ref{fig:SMR}(a) presents the phase map illustrating the CRP and CTP counts as a functions of the driving field parameters $\Omega$ and $\delta$. The number of CRPs depends on the positions of the renormalized transition frequencies $\tilde{\nu}_{\pm}$ relative to the guided-mode band $[\omega_{1},\omega_{2})$, with $\Omega$ serving as the control parameter. As shown in Fig.~\ref{fig:SMR}(d), three distinct regimes emerge: (i) both $\tilde{\nu}_{\pm}$ within $[\omega_{1},\omega_{2})$ yields two CRPs (solid red curve); (ii) one $\tilde{\nu}_{\pm}$ within the band produces a single CRP (dashed green curve); (iii) neither $\tilde{\nu}_{\pm}$ within the band results in no CRPs (dot-dashed blue). These observations are consistent with Figs.~\ref{fig:SMR}(a) and \ref{fig:SMR}(b). In the limit $\Omega\rightarrow0$, the system reduces to a two-level emitter, and the condition for complete reflection simplifies to $\omega-\nu_{-}-\Delta(\omega)=0$, yielding at most one CRP (dotted black curve in Fig.~\ref{fig:SMR}(d)). Furthermore, the $\tilde{\nu}_{\pm}$ in Eq.~(\ref{eq-24}) can also be tuned via the drive detuning $\delta$, while the emitter-waveguide coupling strength $g$ determines the spectral linewidth. As g increases, the Lamb shift increases, broadening the reflection windows and narrowing the transmission windows (see solid red and green dashed curves in Fig.~\ref{fig:SMR}(c)). 

In summary, perfect transmission can be achieved via EIT by adjusting $\delta$, while complete reflection can be induced by tuning $\tilde{\nu}_{\pm}$ through $\Omega$ or $\delta$ to satisfy the Fano resonance condition.

\subsection{Multi-mode regime: input-state-dependent single-photon transport control}
In this subsection, we extend the single-photon transport control from the single-mode to the multi-mode regime. The applicability of single-mode operation is confined to a narrow frequency range $\omega_{1}<\omega<\omega_{2}$, whereas practical applications often demand operation beyond this range. When the photon frequency exceeds $\omega_{2}$, the waveguide  supports multiple transverse modes. The number of active TM modes, $j_{\max}$, is determined by the waveguide geometry and photon frequency, satisfying $\omega_{j_{\max}}<\omega <\omega_{j_{\max}+1}$. This transition to the multi-mode regime reveals two key phenomena: emitter-mediated inter-mode coupling and multi-path quantum interference. The constructive or destructive interference outcomes are determined by the relative amplitudes and phases of indistinguishable scattering paths. With increasing the number of modes, these phenomena collectively dominate the transport dynamics. Eq.~(\ref{eq-17}) indicates that the single-photon scattering in the multi-mode regime is modulated not only  by the external drive through  $G(\omega)$, but also by the incident state through its expansion coefficients {$c_j$}. 

\begin{figure}[ht] 
\centering
\includegraphics[clip=true,width=13cm]{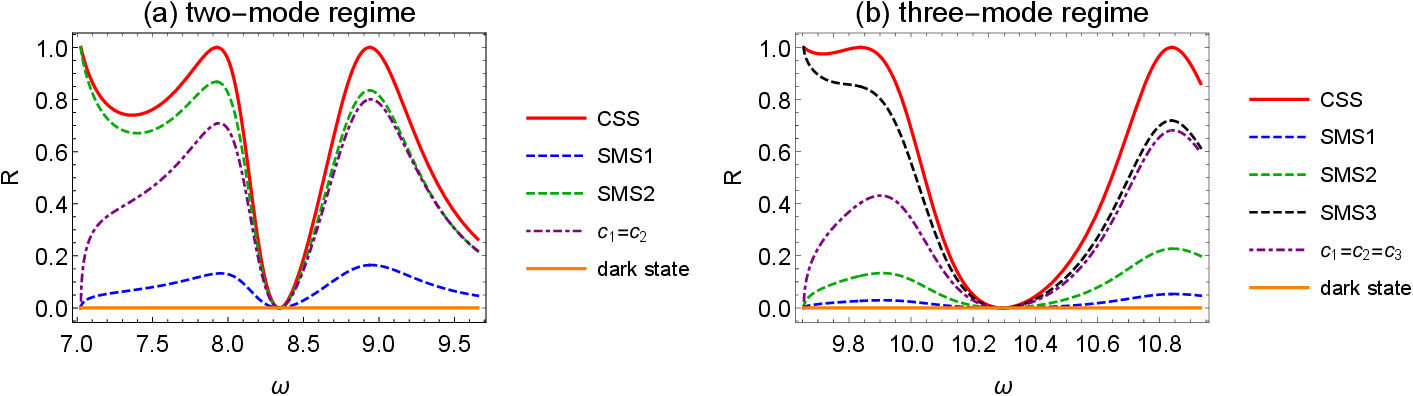}
\caption{Reflectance $R(\omega)$ for different input states in (a) the two-mode and (b) the three-mode regimes. The common parameters are $a=1.5b$, $\Omega=0.5$, and $\delta=0$, with frequencies normalized to $c/b$. (a) Two-mode regime: the results are shown for the specific coherent superposition state (SCSS, $c_{j}\propto\rho_{j}g^{\ast}_{j,k_{j}}$), the single-mode state in the TM$_{11}$ mode (SMS1, $c_{1}=1$) and the TM$_{31}$ mode (SMS2, $c_{2}=1$), the equal-probability superposition state ($c_{1}=c_{2}=1/\sqrt{2}$), and dark state ($\sum_{j=1}^{j_{max}}c_{j}g_{j,k_{j}}=0$). Additional parameters are $g=0.1$, cutoff frequencies $\omega_{2}=7.02$ and $\omega_{3}=9.65$, and $\omega_{e}=(\omega_{2}+\omega_{3})/2$. (b) Three-mode regimes: the results are shown for the SCSS, SMS1, SMS2, the single-mode state in the TM$_{13}$ mode (SMS3, $c_{3}=1$), the equal-probability superposition state ($c_{1}=c_{2}=c_{3}=1/\sqrt{3}$), and dark state. Additional parameters are $g=0.05$, cutoff frequencies $\omega_{3}=9.65$ and $\omega_{4}=10.93$, and $\omega_{e}=(\omega_{3}+\omega_{4})/2$.}
\label{fig:MMR}
\end{figure}

Figure~\ref{fig:MMR} presents the reflection spectra for various input states in the multi-mode regime, focusing on the two- and three-mode regimes under fixed drive parameters, revealing a pronounced input-state dependence. A single-mode state (SMS) injects the photon into only one TM channel, whereas a coherent superposition state distributes it across multiple modes, whose amplitudes interfere through their common emitter interactions. For the coherent superposition state, the scattering amplitude is not a weighted average of individual SMS results. An equal-probability coherent superposition state (purple dot-dashed curve) differs markedly from the arithmetic mean of the SMS curves (dashed). Engineered interference within the coherent superposition state can cancel or enhance all forward waves, yielding unit reflection or transmission even when many channels are open—conditions unreachable by any SMS. The red solid curves, representing the specific coherent superposition state (SCSS, $c_{j}\propto\rho_{j}g^{\ast}_{j,k_{j}}$), reproduce the double-dip perfect reflection signature established in Sec.~III; the orange solid curves (dark state, $\sum_{j}c_{j}g_{j,k_{j}}=0$) show frequency-independent unit transmission; and the dashed SMS curves exhibit only local Fano peaks and never reach $R=1$. This comparison underscores the unique quantum transport control afforded by coherent-superposition-state-mediated interference.

\begin{figure}[ht]
\centering
\includegraphics[clip=true,width=13cm]{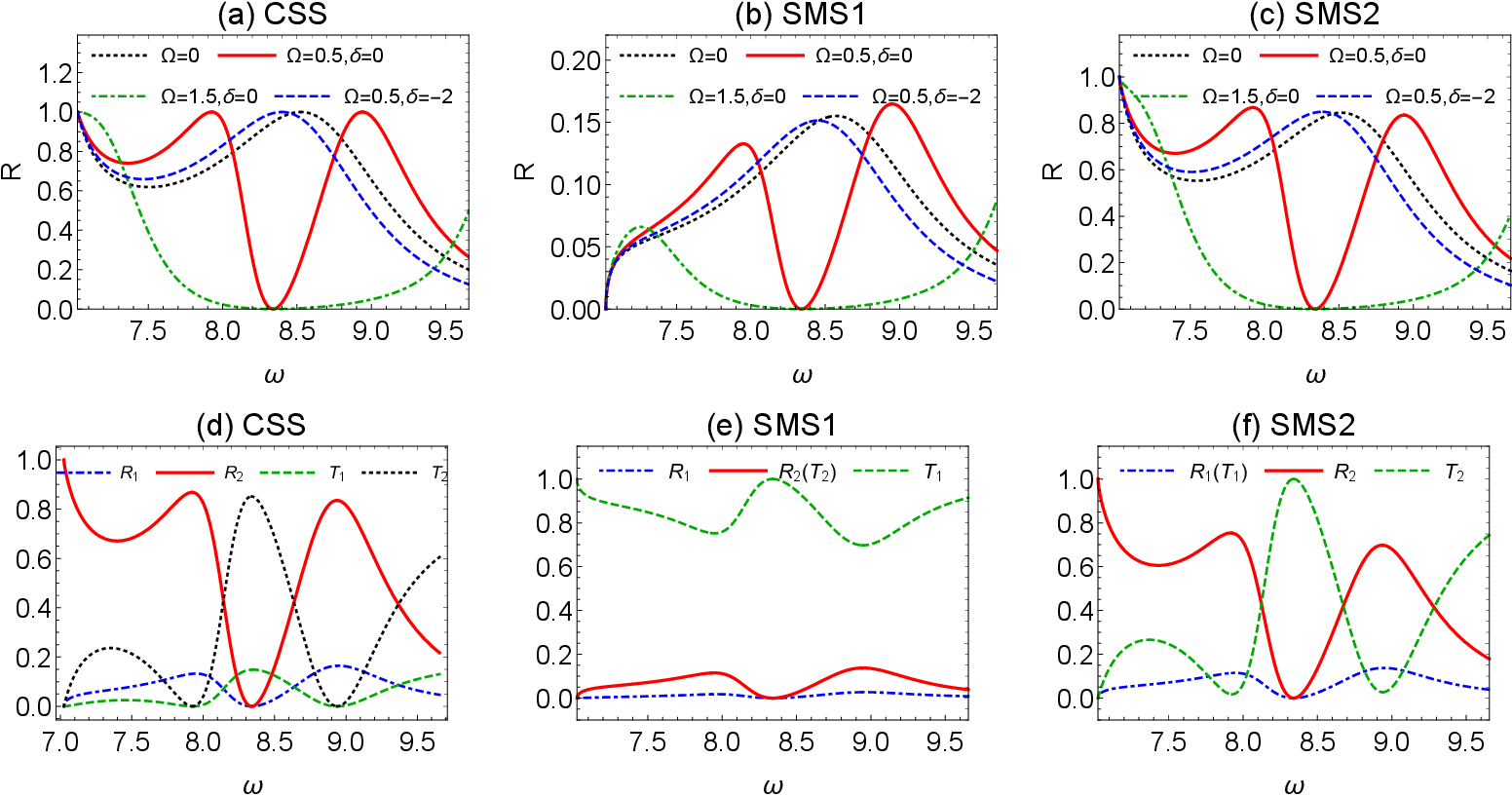}
\caption{Two-mode regime analysis. (a-c) Total reflectance $R(\omega)$ with varying $\Omega$ and $\delta$ for different input photon states: (a) the specific coherent superposition state (SCSS, $c_{j}\propto\rho_{j}g^{\ast}_{j,k_{j}}$), (b) the single-mode state in the TM$_{11}$ mode (SMS1, $c_{1}=1$), and (c) in the TM$_{31}$ mode (SMS2, $c_{2}=1$). (d)-(f) Mode-resolved reflectance $R_{i}(\omega)$ and transmittance $T_{i}(\omega)$ for modes $i=1$ (TM$_{11}$) and $i=2$ (TM$_{31}$) corresponding to the input states in (a-c), respectively, calculated at $\Omega=0.5$ and $\delta=0$. Other common parameters are $a=1.5b$, $g=0.1$, cutoff frequencies $\omega_{2}=7.02$, and $\omega_{3}=9.65$, $\omega_{e}=(\omega_{2}+\omega_{3})/2$, with frequencies normalized to $c/b$.}
\label{fig:TMR}
\end{figure}

We now analyze the single-photon scattering across the multi-mode frequency spectrum by the driven $\Lambda$-type emitter. We systematically compare two incident photon states: the SCSS and the SMS. For the SCSS, we set $c_{j}\propto\rho_{j}g^{\ast}_{j,k_{j}}$ in Eq.~(\ref{eq-16}) to obtain the reflectance and transmittance of the $j$-th mode:
\begin{equation}
R_{j}=\frac{\Gamma_{j}(\omega)\Gamma(\omega)}{\vert G(\omega)\vert^{2}},\quad 
T_{j}=\frac{\Gamma _{j}(\omega)}{\Gamma(\omega)}-R_{j}.\label{eq-26}
\end{equation}
The total reflectance and transmittance are
\begin{equation}
\label{eq-27}
R=1-T=\frac{\Gamma^{2}(\omega)}{\vert G(\omega)\vert^{2}}. 
\end{equation}
A comparison with Eq.~(\ref{eq-25}) reveals that the scattering characteristics of a SCSS resemble those observed in the single-mode regime. Fig.~\ref{fig:TMR}(a) shows the total reflectance $R(\omega)$ across the two-mode regime for varying driving field parameters $\Omega$ and $\delta$. EIT-induced complete transmission, tuned by $\delta$, produces a single CTP at $\omega=\omega_{e}-\delta$ (solid red and dashed green curves). Fano  resonance-induced complete reflection, jointly controlled by $\Omega$ and $\delta$, yields two CRPs when $\text{Re}[G(\omega_{j_{\max}})]<0<\text{Re}[G(\omega_{j_{\max}+1})]$ (solid red curve). Additional control details are listed in Table~\ref{tab:ctp_crp}.  Fig.~\ref{fig:TMR}(d) decomposes the reflectance and transmittance into their modal components to reveal the underlying interference mechanism. Multi-channel interference reduces the transmission amplitude of Eq.~(\ref{eq-15}) to $t_{j}=c_{j}[1-i\Gamma(\omega)/G(\omega)]$, demonstrating that a SCSS input is scattered identically in each TM$_{j}$ mode. In every channel, the direct wave interferes with its scattered counterpart. The dashed green and dotted black curves in Fig.~\ref{fig:TMR}(d) show that Fano resonance drives the transmission amplitude of each mode to zero, resulting in complete reflection, while the solid red and dot-dashed blue curves show that EIT drives the reflection amplitude to zero, yielding complete transmission.

For an input photon in the multi-mode frequency window prepared in the $n$-th SMS $\vert\varphi_{in}\rangle =\vert \varphi_{n,k_{n}}\rangle$, inserting $c_{j}=\delta _{nj}$ into Eq.~(\ref{eq-16}) yields
\begin{equation}
R_{n}=\frac{\Gamma _{n}^{2}(\omega)}{\vert G(\omega)\vert^{2}},\quad
T_{n}=\vert 1-\frac{i\Gamma_{n}(\omega)}{G(\omega)}\vert ^{2},\quad
R_{j}=T_{j}=\frac{\Gamma_{n}(\omega)\Gamma_{j}(\omega)}{\vert G(\omega)\vert^{2}} \ \ (\text{for}\ j\neq n),\label{eq-28}
\end{equation}
with total reflectance and transmittance
\begin{equation}
\label{eq-29}
R=1-T=\frac{\Gamma(\omega )\Gamma_{n}(\omega)}{\vert G(\omega) \vert ^{2}}.
\end{equation}
The emitter absorbs the $n$-th mode photon and decays into all available channels $j$ at rates $\Gamma_{j}$. Figs.~\ref{fig:TMR}(b) and ~\ref{fig:TMR}(c) present the resulting total reflectance in the two-mode regime for SMS1 (TM$_{11}$) and SMS2 (TM$_{31}$) inputs; while Figs.~\ref{fig:TMR}(e) and \ref{fig:TMR}(f) display the corresponding modal components.
Within the input channel ($j=n$), direct and scattered waves interfere partially. At Fano resonance point, this interference drives the transmission to its minimum $T_{n,\min}=[\Gamma(\omega)-\Gamma _{n}(\omega)]^{2}/\Gamma ^{2}(\omega)$ (dashed green curves in Figs.~\ref{fig:TMR}(e) and ~\ref{fig:TMR}(f)), while enhancing the reflection to its maximum $R_{n,\max}=\Gamma^{2}_{n}(\omega)/\Gamma ^{2}(\omega)$ (dot-dashed blue curve in Fig.~~\ref{fig:TMR}(e), solid red curve in Fig.~\ref{fig:TMR}(f)). 
For non-input channels ($j\neq n$), symmetric coupling to left- and right-going waves results in equal reflectance and transmittance $R_{j}=T_{j}$ (solid red curve in Fig.~\ref{fig:TMR}(e) and dot-dashed blue curve in Fig.~ \ref{fig:TMR}(f)), both reaching their peak at the Fano resonance frequency with $T_{j,\max}=R_{j,\max}=\Gamma_{n}(\omega)\Gamma_{j}(\omega)/\Gamma^{2}(\omega)$.
Since $R_{\max}=\Gamma_{n}(\omega)/\Gamma(\omega)<1$, SMS incidence cannot achieve unit reflectance, which is in contrast to the SCSS case. 
Furthermore, only the highest-mode SMS can achieve $R=1$ at its cutoff because $v_{g}=(1/\rho_{j_{\max}})\vert_{\omega=\omega_{j_{\max}}}=0$ (see Fig.~\ref{fig:TMR}(c)), whereas other SMS inputs exhibit full transmission at this frequency (see Fig.~\ref{fig:TMR}(b)). Nevertheless, complete transmission is observed at the EIT point for the SMS case, similar to the SCSS behavior.

Ultimately, single-photon transport in a multi-mode waveguide is dictated by interference arising from coherent superposition states. This leads to perfect reflection for the SCSS and perfect transmission for the dark state. The SMS, conversely, has a limited reflectance peak of $\Gamma_{n}(\omega)/\Gamma(\omega)<1$, only reaching $R=1$ at the highest-mode channel's cutoff frequency. Nonetheless, both state types achieve perfect transmission at the EIT point.

\section{Discussion and conclusion}\label{sec5}
In conclusion, we have presented a comprehensive framework for achieving on-demand control of single-photon transport in broadband, multi-mode waveguides. Our model, featuring a driven $\Lambda$-type emitter coupled to a rectangular waveguide, leverages precisely engineered quantum interference to realize dynamic photon transport manipulation. By employing the Lippmann-Schwinger formalism, we derived the analytical scattering matrix and demonstrated two key interference mechanisms: EIT for complete transmission and Fano resonance for complete reflection. Our scheme facilitates dynamic switching between these mechanisms via the driving field parameters (Rabi frequency $\Omega$ and detuning $\delta$), enabling on-demand control over photon transport. A key capability is the realization of dual-frequency complete reflection, where Fano resonances occur at two tunable frequencies ($\tilde{\nu}_{\pm}$),  which offers remarkable flexibility for spectral engineering.

A central contribution of our work is the revelation of input-state-dependent scattering in the multi-mode regime, a phenomenon without analogue in single-mode physics. We find that generic input states, such as the single-mode state (SMS), cannot achieve perfect reflection due to symmetric scattering into unwanted modes. However, for a specific coherent superposition state (SCSS) with coefficients $c_{j}\propto\rho_{j}g^{\ast}_{j,k_{j}}$, constructive multi-mode quantum interference enables dual-frequency complete reflection. Moreover, the EIT-induced complete transmission is shown to be robust against this modal complexity, ensuring reliable transmission across both regimes.

The proposed system is experimentally feasible across multiple platforms, including optical fibers \cite{Mitsch2014,PhysRevX.5.041036}, photonic crystal waveguides \cite{PhysRevLett.113.113904,Mahmoodian2017}, and superconducting circuits \cite{Bharath2023,PhysRevX.13.021039,PRXQuantum.5.030346}, with the driven $\Lambda$-emitter realizable in systems such as fluxonium qubits \cite{PhysRevLett.120.150504,PhysRevLett.95.087001}, quantum dots \cite{Siampour2023}, or spin defects. Our findings not only advance the understanding of fundamental quantum interference in multi-mode waveguide QED, but also pave the way for developing broadband quantum devices, including dual-frequency filters, multi-mode routers, and on-chip spectrometers. Future directions may explore non-reciprocal scattering, multi-photon dynamics, and many-body effects. This work may establishe a versatile framework for quantum-state-controlled photon routing in broadband quantum networks, with implications for scalable quantum information processing.

\section*{Funding} 
National Natural Science Foundation of China (12205088); Scientific Research Fund of Hunan Provincial Education Department of China (24C0353); Key Laboratory of Opto-electronic Control and Detection Technology of University of Hunan Province (2024HSKFJJ012); Natural Science Foundation of Hunan Province (2025JJ50005).

\section*{Disclosures} 
The authors declare no conflicts of interest.

\section*{Data Availability Statement} 
The data underlying the results presented in this paper are not publicly available at this time but may be obtained from the authors upon reasonable request.

\bibliography{refs}

\end{document}